\documentclass[12pt]{article}
\usepackage{amsmath}
\usepackage{graphicx,psfrag,epsf}
\usepackage{enumerate}
\usepackage{natbib}
\usepackage{amsmath,amsthm,amsfonts,epsfig,amssymb,natbib,eucal,eufrak}
\usepackage{booktabs}
\usepackage{multirow}
\usepackage{siunitx}
\usepackage{qtree}
\usepackage{setspace}

\usepackage{float}
\restylefloat{table}
\usepackage{color}

\usepackage{subcaption}

\newtheorem{thm}{Theorem}

\newtheorem{definition}{Definition}
\newtheorem{ex}{Example}
\newtheorem{result}{Result}
\newtheorem{comment}{Comment}
\newtheorem{remark}{Remark}
\newtheorem{cj}{Conjecture}

\newcommand{\blind}{0}

\begin{document}

\def\spacingset#1{\renewcommand{\baselinestretch}%
{#1}\small\normalsize} \spacingset{1}


\if0\blind
{
  \title{\bf Conjectures on Optimal Nested Generalized Group Testing Algorithm}
  \author{Yaakov Malinovsky
  \\
   Department of Mathematics and Statistics\\ University of Maryland, Baltimore County, Baltimore, MD 21250, USA\\
}
  \maketitle
} \fi

\if1\blind
{
  \bigskip
  \bigskip
  \bigskip
  \begin{center}
    {\LARGE\bf Conjectures on Optimal Nested Generalized Group Testing Algorithm}
\end{center}
  \medskip
} \fi

\bigskip
\begin{abstract}
Consider a finite population of $N$ items, where item $i$ has a probability $p_i$ to be defective.
The goal is to identify all items by means of group testing. This is the generalized group testing problem (hereafter GGTP).
In the case of $\displaystyle p_1=\cdots=p_{N}=p$ \cite{YH1990} proved that
the pairwise testing algorithm is the optimal nested algorithm, with respect to the expected number of tests, for all $N$ if and only if $\displaystyle p \in [1-1/\sqrt{2},\,(3-\sqrt{5})/2]$ (R-range hereafter) (an optimal at the boundary values).
In this note, we present a result that helps to define the generalized pairwise testing algorithm (hereafter GPTA) for the GGTP. We present two conjectures: (1) when all $p_i, i=1,\ldots,N$ belong to the R-range, GPTA is the optimal procedure among nested procedures applied to $p_i$ of nondecreasing order; (2) if all $p_i, i=1,\ldots,N$ belong to the R-range, GPTA the optimal nested procedure, i.e., minimises the expected total number of tests with respect to all possible testing orders in the class of nested procedures.
Although these conjectures are logically reasonable, we were only able to empirically verify the first one up to a particular level of $N$. We also provide a short survey of GGTP.
\end{abstract}

\noindent%
{\it Keywords: Individual testing; pairwise testing}
\medskip

\noindent
{\it AMS Subject Classification:} 05A18, 62P10 \vfill

\newpage
\spacingset{1.45} 
\section{Introduction}
\label{se:I}
\subsection{Common $p$ case}
\noindent
Robert Dorfman introduced the concept of group testing in 1943 as a need to administer syphilis tests to millions of individuals drafted into the U.S. Army during World War II. Interesting historical details related to the problem formulation  can be found in \cite{Dh1999}. The nice description of the \cite{D1943} procedure is given by \cite{F1950}:
 {\it ``A large number, $N$, of people are subject to a blood test. This can be administered in two ways. (i) Each person is tested separately. In this case $N$ tests are required. (ii) The blood samples of $k$ people can be pooled and analyzed together. If the test is negative, this one test suffices for the $k$ people. If the test is positive, each of the $k$ persons must be tested separately, and all $k+1$ tests are required for the $k$ people. Assume the probability $p$ that the test is positive is the same for all  and that people are stochastically independent."
Procedure $(ii)$ is commonly referred to as the Dorfman group testing procedure.

Since then, the group testing has widespread applications. Partial list included quality control in product testing \citep{SG1959}, communication networks \citep{W1985}, American Red Cross screening of blood donations for HIV \citep{D2002}, identification of rare alleles \citep{SAZ2010}, among others.
}

Consider a set $S$ of $N$ items, where each item has the probability $p$ to be defective, and the probability $q=1-p$ to be good independent from the other items. Following the accepted notation in the group testing literature, we call this set a binomial set \citep{SG1959}. A group test applied to the subset $x$ is a binary test with two possible outcomes, positive or negative.
The outcome is negative if all $x$ items are good, whereas the outcome is positive if at least one item among $x$ items is defective. We call such a set defective or contaminated.
The goal is complete identification of all $N$ items with the minimum expected number of tests.

 Every reasonable group testing algorithm should satisfy the following properties \citep{SG1959, U1960}:
(P1) items that are classified as positive or negative will never be tested again, and
(P2) the test is not performed if its outcome can be inferred from previous test results.
In addition, if a subset of good items $I^{\prime}$ is removed from the defective set $I$, then the remaining items $I-I^{\prime}$ form a defective set, and it follows from (P2) that this defective set should not be tested as a whole group.

A {\it nested} class of group testing algorithms was introduced by \cite{SG1959} [see also \cite{H1976} and \cite{YH1990}], and can be described as follows:
\begin{enumerate}
\item[(a)]
At each stage $t$ ($t=0,1,\ldots,T$) of the execution of a nested algorithm, the set $S$ is partitioned into disjoint sets $B_t, C_t$, and $D_t$, where set $C_t$ is a set of classified units, set $B_t$ is a binomial set, and set $D_t$ is a defective set.
At the beginning of the process at stage (stage 0), $B_{0}=S$, and both $C_{0}$ and $D_{0}$ are empty. At the termination of the process (stage $T$), $C_{T}=S$, and both $B_{T}$ and $D_{T}$ are empty.
If at any stage during the process $|D_{t}|=1$, then, according to (P1) above, this sole defective item should be moved from set $D_{t}$ into set $C_{t}$.
\item[(b)]
At each stage $t$ of the algorithm execution, if $D_{t-1}$ is not empty, then a proper subset $D^{\prime}_{t-1}$ of $D_{t-1}$ is tested. If the outcome of tesing $D^{\prime}_{t-1}$ is positive, then $C_t=C_{t-1}$, $D_{t}=D^{\prime}_{t-1}$ and $B_t=N-C_{t}-D_{t}$ (follows from Result 1 below); if the outcome of testing $D^{\prime}_{t-1}$ is negative, then $C_{t}=C_{t-1}+D^{\prime}_{t-1}$, $D_t=D_{t-1}-D^{\prime}_{t-1}$, and $B_{t}=B_{t-1}$.
Otherwise, if $D_{t-1}$ is empty and $B_{t-1}$ is not empty, then a subset $B^{\prime}_{t-1}$ of $B_{t-1}$ is tested.
If the outcome of testing $B^{\prime}_{t-1}$ is positive, then $C_t=C_{t-1}, D_{t}=B^{\prime}_{t-1}$, and $B_{t}=B_{t-1}-B^{\prime}_{t-1}$;
if the outcome of testing $B^{\prime}_{t-1}$ is negative, then $C_t=C_{t-1}+B^{\prime}_{t-1}, D_{t}=D_{t-1}$, and $B_{t}=B_{t-1}-B^{\prime}_{t-1}$.
\end{enumerate}

An optimal nested procedure in the form of a dynamic programming algorithm was found by \cite{SG1959}.
Subsequently, \cite{S1960} and \cite{H1976} improved its computational efficiency. Recently, \cite{ZP2016} provided an asymptotic analysis of the optimal nested procedure.
Finally, different aspects concerning the nested class of group testing
procedures were summarized and investigated in \cite{MA2019}.
For $N=2$, the optimal algorithm coincides with Huffman's \citep{H1952} encoding algorithm \citep{S1967}. However, the optimal nested algorithm is not optimal for $N\geq3$ \citep{S1960,S1967}.
An explicit construction, an example showing that the optimal nested procedure is not in fact optimal, can be found in Section 13 of \cite{S1960}.

Until today, an optimal group testing procedure for complete identification under a binomial model is unknown for $\displaystyle p<(3-\sqrt{5})/2$ and general $N$.  For $\displaystyle p\geq(3-\sqrt{5})/2$ (or $3-q-q^2\geq 2, q=1-p$) \cite{U1960} proved that the optimal group testing procedure is individual, one-by-one testing (at the boundary point it is an optimal).

The pairwise nested algorithm belongs to the nested class and  was defined by \cite{YH1990}.
A verbatim definition of it is as follows:
\smallskip

{\it
We define the pairwise testing algorithm by the following two rules:
\begin{itemize}
\item[(i)]
If no contaminated set exists, then always test a pair from the binomial set
unless only one item is left, in which case we test that item.
\item[(ii)]
If a contaminated pair is found, test one item of that pair. If that item is
good, we deduce the other is defective. Thus we classify both items and only a
binomial set remains to be classified. If the tested item is defective, then by a result
of \cite{SG1959}, the other item together with the remaining binomial set
forms a new binomial set. So, both cases reduce to a binomial set.
It is easily verified that at all times the unclassified items belong to either a
binomial set or, a contaminated pair. Thus the pairwise testing algorithm is well
defined and is nested.
\end{itemize}

}

The following result offers a closed-form design for the optimal nested procedure, which can be resolved without
computational effort, provided that all $p_i, i=1,\ldots,N$ belong to the R-range.

\begin{thm}{\bf \cite{YH1990}}\\
\label{re:YH}
The pairwise testing algorithm is the unique (up to the substitution of
equivalent items) optimal nested algorithm for all $N$ if and only if $1-1/\sqrt{2}\leq p \leq (3-\sqrt{5})/2$(at the boundary values the pairwise testing algorithm is an optimal
nested algorithm).
\end{thm}

\subsection{The generalized group testing problem}

The generalized group testing problem (GGTP): $N$ stochastically independent units $u_1,u_2,\ldots,u_N$, where unit $u_i$ has the probability $p_i$ ($0<p_i<1$) to be defective and the probability $q_i=1-p_i$ to be good. We assume that the probabilities $p_1, p_2,\ldots,p_N$ are known and we can decide the order in which the units will be tested. All units have to be classified as good or defective by group testing.
The generalized group testing problem was first introduced by \cite{S1960} on page 144. In this work, two (or more) different kinds of units are presented and can be put into the same test group. In the case of two kinds of units with known probabilities $q_1\geq q_2$, the individual testing is optimal if $3-q_1-q_1q_2>2$.
This result follows the \cite{H1952} encoding algorithm construction when $N=2$ \citep{S1960}.
Since its introduction, GGTP has been investigated (\cite{LS1972, NS1973, K1973, N1975, H1976, YH1988a, YH1988b, KS1988, YH1990}; \cite{KJP2014}; \cite{M2017}).
Even for a particular nested group testing algorithm the optimal regime (or, order in which groups/units will be tested ) is known only for
for the Dorfman procedure \citep{D1943} because of \cite{H1976}.

For the GGTP, \cite{H1976} proved that under Dorfman's procedure an optimal partition is an ordered partition  (i.e., each pair of subsets has the property such that the numbers in one subset are all greater or equal to every number in the other subset).  Then Dorfman's procedure
is performed on each subset.
It allowed Hwang to find the optimal solution
using a dynamic programming algorithm with the computational effort $O(N^2)$.
But, even using a slightly modified Dorfman procedure or \cite{S1957} procedure, the ordered partition is not optimal \citep{M2017}.
 As the total number of possible partitions
is the Bell number, it is impossible to use brutal search to obtain an optimal solution, which is unknown \citep{H1981, M2017}.
\cite{KS1988} provided a dynamic programming (DP) algorithm having computational effort $O(N^3)$ to find an optimal nested procedure for a given order of units $u_1,\ldots,u_N$ (which order should be preserved at all stages of the testing process).
In addition, \cite{KS1988} used the \cite{U1960} method and extended \cite{S1960} result from $N=2$ to general $N$. Namely, they proved
that if $3-q_1-q_1q_2>2$, where $q_1\geq\cdots \geq q_N$, then individual testing is optimal.
Closely related results were obtained by \cite{YH1988b}, and can be summarized as follows:
\begin{thm}{\bf \cite{YH1988b}}\\
\label{re:YH88}
Assume without loss of generality that $0<p_1\leq p_2\leq \cdots \leq  p_N<1$. Then,
\begin{enumerate}
\item
If $3-q_1-q_1q_i>2$, then there exists an optimal algorithm which tests $u_i$ individually.
\item
Denote by $k=\sup\Big\{\sup\{i=1,\ldots,N:3-q_1-q_1q_i>2\},0\Big\}$, with $\sup\{\phi\}=-\infty$.
Then there exists an optimal algorithm which tests $u_{k+1},\ldots,u_N$ individually.
\item
If there exists an optimal algorithm in which $u_i$ is tested individually, there exists an optimal algorithm in which $u_j$ is tested individually for all $j$ with $p_j>p_i$.
\end{enumerate}
\end{thm}

It is important to note that in contrast to \cite{U1960}, the results by \cite{KS1988}  and \cite{YH1988b}  provide a sufficient, but not necessary, condition.
\cite{YH1988b} constructed an example with $N=3, p_1=0.1, p_2=0.2, p_3=0.8$ such that $3-q_1-q_1q_3<2$, where the optimal algorithm tests $u_3$ individually.
In contrast, if $p_i<(3-\sqrt{5})/2$ for every $i=1,\ldots,N$, then $3-q_j-q_jq_i<2$ for every $i,j=1,\ldots,N$ and therefore no item should be tested individually unless there are no items left to combine.
In addition, it was shown in \cite{YH1988a} that $E^{*}\left(p_1,\ldots,p_{N}\right)$ is nondecreasing in each $p_i<1$ for every $N$, where $E^{*}\left(p_1,\ldots,p_{N}\right)$ denotes the expected number of tests for an optimal algorithm in GGTP.
In combination with \cite{U1960}, this result implies that if $p_i>(3-\sqrt{5})/2$ for every $i=1,\ldots,N$, then the optimal group testing procedure is to perform individual, one-by-one testing.

\section{Description of the Problem, Results and Examples}
We want to define the generalized pairwise testing algorithm (GPTA) for the GGTP. Two results below will help to proceed.
The first result is a simple generalization of \cite{SG1959} result for the common $p$ case into GGTP (see also \cite{KS1988}).
\begin{result}[\cite{SG1959}]
\label{re:1}
In the GGTP, given a defective set $I$ and given that a proper subset $I_1,\,I_1\subset I$ contains at least one defective unit, then the posteriori
distribution of the units in the subset $I-I_1$ is the same as it was before any testing.
\end{result}
The second result describes an optimal rule for nested testing in the case that at some stage, we have to test two particular units $a$ and $b$.
\begin{result}
\label{re:Order}
Suppose that a nested procedure is applied.
Also suppose that the $n$ units that remain to be tested, $a,b,u_3,\ldots,u_n$, all have unknown status, and the corresponding  probabilities of those units being good are $q_a, q_b, q_3,\ldots, q_n$. We start by testing two units together, as a group, with the corresponding probabilities $q_a$ and $q_b$, where $q_a\geq q _b$. Then, under this setting,
when the first group test of units $a$ and $b$ is positive, we then have to test the unit for which the corresponding probability of being good is largest,
i.e. unit $a$ (call it algorithm A).
If the outcome of testing unit $a$ is negative, then the second unit is positive by deduction.
Otherwise, if the outcome of testing unit $a$ is positive, then by Result \ref{re:1} the conditional distribution of the status of the second unit is a Bernoulli distribution with parameter $p_b=1-q_b$, and units $b,u_3,\ldots,u_n$ remain to be tested.
\end{result}

\begin{proof}
The proof is based on direct comparison of two possible algorithms, namely, algorithms A and B,
where, in algorithm B, we first test unit $b$ individually.
Denote $T$ as the total number of tests and denote $\displaystyle E\left(p_{i_{1}},\ldots,p_{i_{k}}\right)$ as the total expected number of tests of units $i_1,\ldots,i_k$ with the corresponding probabilities  $\displaystyle p_{i_{1}},\ldots, p_{i_{k}}$ under a nested procedure.  The left branch of the tree below represents a negative test result, and the right branch represents a positive test result.
\bigskip

\begin{figure}[h]
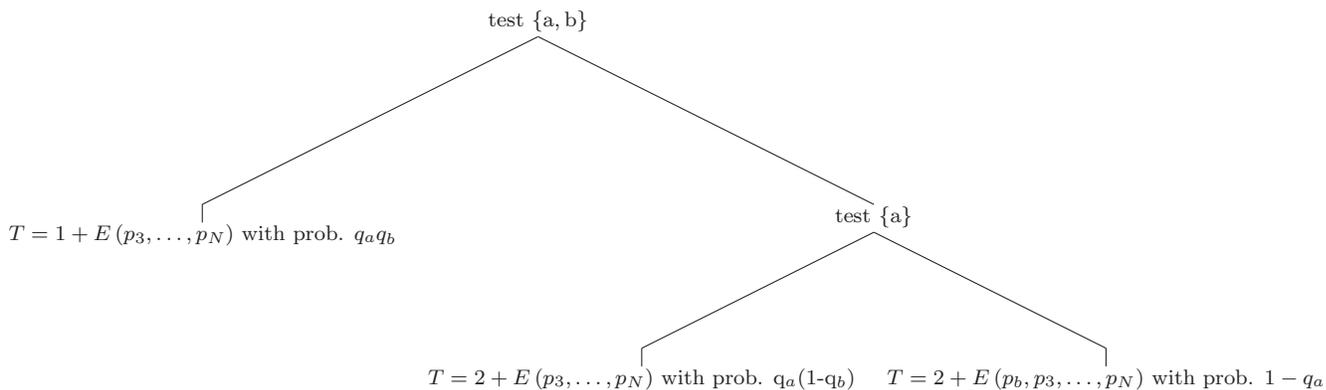

        \scriptsize
        \Tree [.{test \{a,\,b\}} [{$T=1+E\left(p_3,\ldots,p_N\right)$ with prob. $q_a q_b$} ] [.{test \{a\}} [{$T=2+E\left(p_3,\ldots,p_N\right)$ with prob. q_a(1-q_b)} ] [{$T=2+E\left(p_b, p_3,\ldots,p_N\right)$ with prob. $1-q_a$}  ] ] ]
        \caption{Algorithm A}
        \label{fig:A}
\end{figure}
\bigskip

\begin{figure}[h]
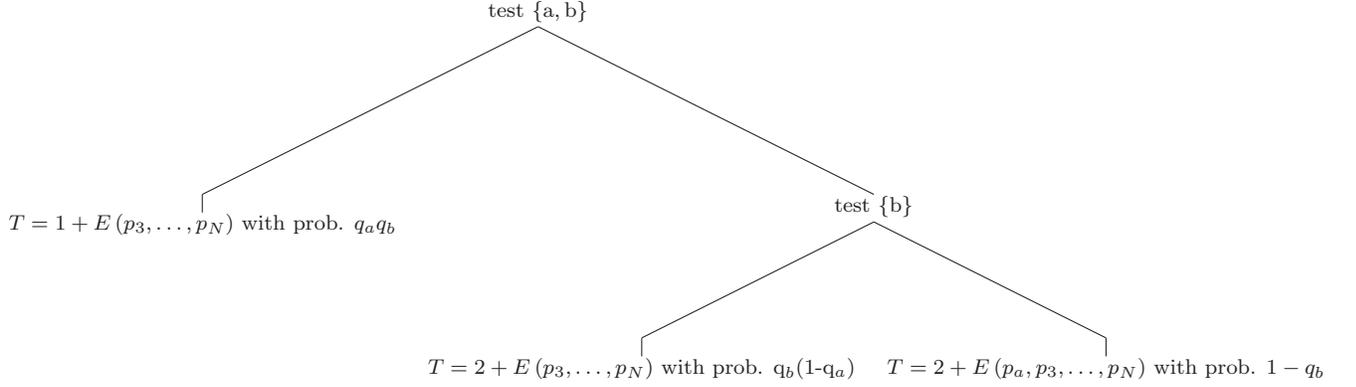

        \scriptsize
        \Tree [.{test \{a,\,b\}} [{$T=1+E\left(p_3,\ldots,p_N\right)$ with prob. $q_a q_b$} ] [.{test \{b\}} [{$T=2+E\left(p_3,\ldots,p_N\right)$ with prob. q_b(1-q_a)} ] [{$T=2+E\left(p_a, p_3,\ldots,p_N\right)$ with prob. $1-q_b$}  ] ] ]
        \caption{Algorithm B}
        \label{fig:B}
\end{figure}

Let $E_{A}(T)$ and $E_{B}(T)$ be the expected total number of tests under algorithms $A$ and $B$ correspondingly.
We have,
\begin{align*}
&
E_{A}(T)=q_a E\left(p_3,\ldots,p_N\right) +(1-q_a)E\left(p_a, p_3,\ldots,p_N\right)+2-q_a q_b.\\
&
E_{B}(T)=q_b E\left(p_3,\ldots,p_N\right) +(1-q_b)E\left(p_b, p_3,\ldots,p_N\right)+2-q_a q_b.
\end{align*}
Since $\displaystyle E\left(p_1,p_3,\ldots,p_k\right)$ is non-decreasing in each $p_i$ for $0\leq p_i\leq 1$ \citep{YH1988a} and we assume w.l.g. that
$p_a\leq p_b$,
we have $\displaystyle E\left(p_a, p_3,\ldots,p_N\right)\leq E\left(p_b, p_3,\ldots,p_N\right)$. Therefore, we obtain
\begin{align*}
&
E_{A}(T)-E_{B}(T)\leq \left(q_b-q_a\right)\left(E\left(p_b, p_3,\ldots,p_N\right)-E\left( p_3,\ldots,p_N\right)\right)\leq 0.
\end{align*}

The last inequality follows from the obvious fact that $\displaystyle E\left( p_3,\ldots,p_N\right)\leq E\left(p_b, p_3,\ldots,p_N\right)$.
\end{proof}

\begin{remark}
The intuition behind Result \ref{re:Order} is as follows: Suppose that the pair $\left\{a, b\right\}$ test is positive. Then, if a subsequent individual test of one unit from the set $\left\{a,b\right\}$ is negative, we can conclude by deduction (without actual testing) that the second unit is positive (possibility 1). Alternately, if the subsequent individual test is positive, then the status of the remaining unit is unknown, and this unit will at some stage be tested in the group or
individually (possibility 2). Since $q_a\geq q_b$ and we prefer possibility 1
over possibility 2, we should select unit $a$ to be tested first.
\end{remark}

Now we are ready to define GPTA for GGTP.

\begin{definition}
\label{def:pair}
Let $u_1,u_2\ldots,u_N$ be the fixed initial order of units to test, for which
the corresponding probabilities of being good are $q_1,\ldots, q_N$.
We define the generalized {\it pairwise} testing algorithm (GPTA) by the following rules:
\begin{itemize}
\item[(a)]
Test the pair $\{u_1,u_2\}$. If the outcome is negative, then continue by testing the next pair unless
only one unit is left, in which case we test that unit.
\item[(b)]
If the outcome is positive, then test the unit with the greater probability of being good,
i.e. unit $u_{j_1}$ where $j_1=\displaystyle \arg \max(q_1, q_2)$.
If unit $u_{j_1}$ is found to be good, then the other unit $u_{j_2}$, where $j_2=\displaystyle \arg \min(q_1, q_2)$, is defective by deduction.
Otherwise, if the tested unit $u_{j_1}$ is defective,
then by Result \ref{re:1} the conditional distribution of the status of $u_{j_2}$ is a Bernoulli distribution with parameter $p_{j_2}=1-q_{j_2}$, and units $u_{j_2},u_3,\ldots,u_n$ remain to be tested.
Continue with testing the next pair of units.

\end{itemize}
\end{definition}

Note that GPTA does not necessarily preserve the initial predetermined testing order; i.e. even if a defective unit $u_i$ is tested no later than unit $u_j$, $u_i$ may remain in the testing process even after unit $u_j$ is identified. However, if the initial predetermined testing order follows a nondecreasing order of $p_i$s ($p_1\leq p_2\leq \cdots \leq p_N$), then GPTA preserves the initial testing order.

It is natural to expect that the result of \cite{YH1990} will hold for GGTP in the case of $\displaystyle 1-1/\sqrt{2}\leq p_i \leq (3-\sqrt{5})/2, \,\,i=1,\ldots,N.$
The following example helps us to understand this situation. Here, we compare GGTP with an optimal nested procedure from \cite{KS1988} for all possible testing orders.
Their procedure requires the initial testing order to be preserved throughout the testing process; otherwise its computational complexity will be exponential as a function of $N$. Therefore, the procedure of \cite{KS1988} does not necessarily satisfy the optimal rule obtained in Result \ref{re:Order} in the case where testing two units,
unless the $p_i$s are arranged in nondecreasing order.

\begin{ex}
\label{ex1}
Suppose $\left\{q_1,\,q_2,\,q_3,\,q_4\right\}=\left\{  0.62,\,\,\,\,0.62,\,\,\,\,\,0.65,\,\,\,\,0.68\right\}$.

\begin{table}[H]
\caption{The expected total number of tests $E_{P}(T)$ for all possible initial testing orders under the GPTA and
the expected total number of tests $E_{Ne}(T)$ under an optimal nested ordered procedure following the algorithm by \cite{KS1988}.}
\begin{center}
\begin{tabular}{lllll|llll}
\toprule
Permutation& \multicolumn{4}{c}{Initial Testing Order} & $E_{P}(T)$ & $E_{Ne}(T)$\\
\midrule
{1}& {0.68}&{0.65}&{0.62}&{0.62}&  {3.8576}&{3.8576} \\
2& 0.68&0.62&0.65&0.62&{\bf3.8449} &3.8454\\
3&0.68&0.62&0.62&0.65& 3.8545  &3.8754\\
4&0.65&0.68&0.62&0.62& 3.8576  &3.8691\\
5&0.65&0.62&0.68&0.62&{\bf 3.8449}  &3.8454\\
6&0.65&0.62&0.62&0.68& 3.8659  &3.9054\\
7&0.62&0.65&0.68&0.62& {\bf 3.8449}  &3.8655\\
8&0.62&0.65&0.62&0.68& 3.8659  &3.9255\\
9&0.62&0.68&0.65&0.62&{\bf 3.8449}  &3.8610\\
10&0.62&0.68&0.62&0.65&3.8545  &3.8910\\
11&0.62&0.62&0.68&0.65&3.8749  &3.8736\\
12&0.62&0.62&0.65&0.68&3.8863  &3.9036\\
\bottomrule\\
\end{tabular}
\end{center}
\end{table}

\end{ex}

\begin{comment}(Example \ref{ex1})
The following observations were made:
\begin{itemize}
\item[(a)]
For the initial ordered testing $q_1\geq q_2 \geq q_3 \geq q_{4}$ (permutation 1) both algorithms are identical.
This permutation is not optimal.
\item[(b)]
In all cases, instead of permutation 1, the procedure by \cite{KS1988} for the given order differs from GPTA, but in all cases the testing group size under their  procedure does not exceed 2 and can be 1.
For example, under permutation 2 this procedure is presented below with the corresponding $E_{Ne}=3.8454$.
\begin{center}
\Tree [.{test (0.68, 0.62)} [{test (0.65, 0.62)}  ] [.{test (0.68)} [{test (0.65, 0.62)} ] [{test (0.62) + test (0.65, 0.62) }  ] ] ]
\end{center}
\item[(c)]
For the permutation 11 the GPTA is not optimal.
\item[(d)]
The expected length of an optimal prefix Huffman code, which serves as a theoretical and generally non-attainable lower bound \citep{NS1973}, is 3.7977344.
\item[(e)]
Initial testing orders 2,5,7, and 9 are optimal with the corresponding expected number of tests equals $E_{P}(T)=3.8454$. GPTA corresponding to an initial testing order 2 is presented below.
\begin{center}
\footnotesize
\qtreecenterfalse
{{
\Tree [.{test $\left\{0.68, 0.62\right\}$}
[.{test $\left\{0.65, 0.62\right\}$}
{Stop} [.{test $\left\{0.65\right\}$} {Stop} {test $\left\{0.62\right\}$}
]
]
[.{\,\,\,\,\,\,test $\left\{0.68\right\}$}
[.{test $\left\{0.65, 0.62\right\}$}
{Stop} [.{test $\left\{0.65\right\}$} {Stop} {test $\left\{0.62\right\}$}
]]
[.{test $\left\{0.62, 0.65\right\}$} {test $\left\{0.62\right\}$}
[.{test $\left\{0.65\right\}$} {test $\left\{0.62\right\}$}
[.{test $\left\{0.62, 0.62\right\}$} {Stop}
[.{test $\left\{0.62\right\}$} {Stop} {test $\left\{0.62\right\}$}
]
]
]
]
]
]
}}
\end{center}

\end{itemize}
\end{comment}

We conjecture that the result of \cite{YH1990} (Theorem \ref{re:YH}) holds for GGTP. That is, for a given testing order concerning the values $p_1 ,\ldots,p_N$, an optimal design in the closed form can be determined without any computational effort. The precise formulation of this conjecture is presented in the next section.

\section{Conjectures}

\begin{cj}
\label{cj:1}
Given that $u_1,u_2\ldots,u_N$ are labeled according to a non-decreasing order of $p_1\leq p_2\leq \cdots \leq p_N$,
such that $1-1/\sqrt{2}\leq p_i\leq (3-\sqrt{5})/2$ for $i=1,\ldots,N$,
GPTA is the optimal nested ordered algorithm (at the boundary values, the pairwise testing algorithm is an optimal
nested algorithm).
\end{cj}

Conjecture \ref{cj:1} was empirically verified for $N\leq 1000$ in the following manner:
We generated $N$ values from a continuous uniform $[1-1/\sqrt{2},\, (3-\sqrt{5})/2]$ distribution and ordered them such that $p_1\leq p_2\leq \cdots p_N$ . Then, we applied the optimal ordered (with respect to $u_1,\ldots,u_N$)
nested procedure by \cite{KS1988} along with the optimal pairwise testing procedure.
For this particular order, the optimal rule presented in Result 2 automatically holds for the algorithm of \cite{KS1988}.
In both procedures, the expected total number of tests was calculated to verify that the difference between those expectations equals zero. We repeated this process a number of times. However, since the computational effort of the \cite{KS1988} algorithm is proportional to $N^3$, it is not computationally feasible to make many repetitions when $N$ is large.
Therefore, the number of repetitions was chosen as a decreasing function of $N$.
For the first 100 smallest values of $N$, we repeated the process 500 times; and for each successive 100 values of $N$, we decreased the number of repetitions by half,
ultimately performing only a single repeat for the 100 largest values of $N$.

\begin{cj}
\label{cj:2}
For all integer positive values of $N$ and all $p_i, i=1,\ldots,N$ in the interval $[1-1/\sqrt{2}, (3-\sqrt{5})/2]$, the generalized pairwise testing algorithm is the optimal nested procedure (at the boundary values, GPTA is an optimal nested algorithm); that is, within the class of nested procedures, this approach
minimizes the expected total number of tests with respect to all possible testing orders.
\end{cj}

\begin{remark}
For $N=2$ and $1-1/\sqrt{2}\leq p_i \leq (3-\sqrt{5})/2,\,\,i=1,2$,  the optimal nested algorithm is GPTA and it is also the optimal group testing procedure because it coincides with Huffman's \citep{H1952} encoding algorithm.
\end{remark}

If Conjecture \ref{cj:2} is true, it is not clear whether the problem of finding the optimal GPTA
with respect to all possible testing orders is a computational tractable problem \citep{GJ1979}.
But, it still may be possible to provide proof of existence.

It was suggested by an anonymous reviewer that for even values of $N$, a guess for an optimal ordering of items may be as follows: Split the units $u_1,u_2,\ldots,u_N$ with corresponding probabilities $p_1\leq p_2\leq \cdots \leq p_N$ into subsets $\underline{U}=\left\{u_1,\ldots, u_M\right\}$ and $\overline{U}=\left\{u_{M+1},\ldots, u_N\right\}$. Then apply GPTA to the testing order $u_1,u_{M+1},u_2,u_{M+2},\ldots,u_{M},u_{N}$. This order appears to be optimal for GPTA in the case of $N=4$, as that was empirically verified. However, this ordering is not optimal for the next even value of $N$, i.e. $N=6$.
At this stage, we do not have a good guess for a best ordering.

\paragraph{Acknowledgement.}
The author thanks the associate editor and two anonymous reviewers for their exceptionally insightful and
helpful reports, which led to significant improvements in the paper.

{}

\end{document}